\documentstyle[12pt,epsf]{article}
\newcommand{\spone}{1.1}

\newcommand{\singlespace}{\edef\baselinestretch{\spone}\Large\normalsize}

\title{A Self Assembled Nanoelectronic Quantum Computer
Based on the Rashba Effect in Quantum Dots}
\author{S. Bandyopadhyay$\thanks{Corresponding author. E-mail: bandy@quantum1.unl.edu}$\\
\\Department of Electrical Engineering \\
University of Nebraska \\
Lincoln, Nebraska 68588-0511, USA}
\date{}

\begin{document}

\maketitle

\singlespace

\begin{abstract}

Quantum computers promise vastly enhanced computational power and an
 uncanny ability to solve classically intractable problems. However, few proposals exist for robust, solid state implementation of such computers
where the quantum gates are sufficiently miniaturized to have nanometer-scale
dimensions. Here I present a new design for a nanoscale universal
quantum gate. It
consists of two adjacent quantum dots each containing a single electron. 
The spin of each electron encodes a qubit. Ferromagnetic contacts
are used to coherently inject an electron into each dot with a 
definite spin orientation, thus defining the initial state of the qubit. To rotate the qubit, we exploit the fact that the ground state of the electron  is spin-split because of the magnetic field 
caused by the ferromagnetic contacts, as well as the Rashba interaction
arising from spin-orbit coupling.
Arbitrary qubit rotations are effected by modulating the Rashba splitting with an external electrostatic potential and bringing the total spin splitting
energy in a target quantum dot in resonance with a global ac magnetic 
field. . The controlled dynamics of the universal
2-qubit rotation operation can be realized by controlling the exchange
coupling between the two dots with yet another gate potential
which changes the overlap between the wavefunctions of the two electrons. The qubit (spin orientation) is
read via the current induced between the ferromagnetic layers under 
an applied potential. The ferromagnetic layers act as ``polarizers''
and ``analyzers'' for spin injection and detection. A complete 
prescription for initialization of the computer and data input/output
operations is presented. Finally, we define a clear pathway towards 
self assembling this structure using chemical synthesis and some 
lithography.

\end{abstract}


\pagebreak

\section{Introduction}

There is significant current interest in quantum computers because they possess vastly enhanced 
capabilities accruing from quantum parallelism \cite{schumacher,steane}.
Some quantum computing algorithms \cite{shor,grover} have been shown to be able to solve classically intractable 
problems. i.e. perform tasks that no classical 
algorithm could perform efficiently or tractably. Quantum computers and 
quantum memories are also the backbone of quantum repeaters and 
teleportation networks

Experimental effort in realizing quantum computers has been
geared towards  synthesizing  universal
quantum logic gates from which  quantum computers can be built. A universal gate is a 
2 qubit-gate \cite{divincenzo,Lloyd,barenco} and has basically two 
attributes. First, it allows arbitrary 
unitary rotations on each qubit and second, it performs the quantum controlled 
rotation operation whereby one of the qubits (the target qubit)
is rotated through an arbitrary angle, if, and only if, the other
qubit (the control qubit) is oriented in a specified direction.
The orientation of the control qubit is left unchanged. It is this
conditional dynamics of the controlled rotation operation that is challenging
to implement experimentally.

Recently, it has been shown \cite{shor1,kitaev,roychowdhury} that there
exist universal falut-tolerant computers that can operate in a non-ideal
noisy environment. They are usually a circuit composed of one- or
two-qubit gates performing various unitary rotations on a qubit
(e.g Hadamard, Pauli rotations through rational or
irrational angles, etc.). They too can, in principle, be realized from the 
basic gate that we discuss here.

The most vexing  problem in experimentally realizing quantum computers is  the issue of 
decoherence. Qubits are coherent superpositions of two-level states
and, as such, are  delicate entities. 
Any coupling to the environment will destroy the coherence of the superposition state and corrupt the
qubit.
Were it not for the recent discovery of quantum error correcting 
codes \cite{calderbank} that can correct errors due to decoherence
through the use of appropriate {\it software}, quantum computing 
would have  remained a theoretical curiosity.

In the past, atomic systems were proposed as ideal testbeds for experimental
quantum logic gates because of the relatively long coherence times associated 
with the quantum states of trapped atoms and ions \cite{cirac}.
Experimental demonstrations of quantum logic gates 
were carried out in atomic systems \cite{turchette,monroe}.
Recently, nuclear magnetic resonance (NMR) spectroscopy has been shown to be an 
attractive alternative \cite{cory,gershenfeld} and there has been 
some reports of experimental demonstrations involving NMR \cite{jones}. However, there are also some doubts regarding the efficacy of NMR based approaches when dealing
with many qubits \cite{braunstein}.

The major drawback of both atomic and NMR systems is of course that  they are unwieldy, expensive and inconvenient. Solid state (especially nanoelectronic)
implementations would be much more desirable because they are 
amenable to miniaturization. One would like a quantum gate that is
one nanometer long and not one meter long. The technology base
that exists in the solid-state area with regards to  miniaturization
is unparalleled.

While it is understood that solid state systems will be preferable
vehicles for quantum computation,  it is also well-known that the phase memory 
time of charge carriers in solids saturate to only a few nanoseconds
as the lattice or carrier  temperature is  
lowered to a few millikelvins \cite{mohanty} (this is caused by coupling 
of carriers to the zero-point motion of phonons). Thus, solid state 
implementations of quantum gates where the qubits are coupled to
charge degrees of freedom will be always dogged by serious decoherence 
problems. Even though such systems have been proposed in the past
\cite{quantum-wells,bandy,balandin}, they will require clock speeds
in the far infra-red frequency range to meet 
Preskill's criterion for fault-tolerant computing \cite{preskill}.
 
A possible solution of this problem is to use the spin degrees of 
freedom in solid state systems to encode qubits since the spin is 
not coupled to 
electromagnetic noise and hence should have much longer coherence
times than charge. It has been shown that electronic and nuclear spins  of phosphorus dopant
atoms $^{31}$P in silicon have very long spin-flip times 
(or the co-called $T_1$ times in the language of spectroscopy) of about an 
hour \cite{faher}. Consequently, nuclear spins of $^{31}$P dopant atoms in silicon have been advocated as preferable vehicles for qubits \cite{privman,kane,vrijen}.
However, the actual coherence time (or $T_2$ time) 
of electron spin in P-doped silicon may be on the order of a
millisecond. Compound semiconductors may exhibit somewhat shorter 
spin coherence times, but spin coherence times 
as long as 100 ns have been experimentally demonstrated in
n-type GaAs at the relatively balmy temperature of 5 K \cite{kikkawa}. 
Thus, it is practical to contemplate solid state
quantum computers based on single electron spins.

Not all semiconductors however
are suitable hosts for qubits. Pyroelectric 
materials (uniaxial crystals without inversion symmetry) 
usually exhibit electric dipole spin resonance which can increase 
the spin flip rate significantly \cite{romestain} by strongly coupling the
spin to phonons. An advantage of quantum dots is that the spin-phonon interaction
may be reduced because of a constriction of the 
phase space for scattering. Moreover, the phonon-bottleneck
effect  \cite{benisty}
may block phonon-induced spin-flip transitions. Another obvious strategy to
increase the coherence time is
to decrease the phonon population by reducing the temperature. The temperature  must be low in any case since 
the time to complete a quantum calculation should not significantly exceed
the thermal time scale $\hbar/kT$ \cite{unruh} irrespective of any 
other consideration. 

Quantum gates based on spin polarized single electrons housed in quantum 
dots have been proposed by us in the past \cite{bandy1} and more 
recently by Loss and DiVincenzo \cite{loss}. Here we adopt a different
idea - which is still based on spin-polarized single electrons - to provide 
a realistic paradigm for the realization of a {\it self-assembled}
solid-state, nanoelectronic universal quantum gate.

\section{A Self Assembled Universal Quantum Gate} 

Consider two adjacent {\it penta-layered} quantum dot structures shown in Fig. 1(a).
In each dot, the two outer layers are ferromagnetic and the middle layer is a 
semiconductor. Insulating layers separate the ferromagnetic
layers from the semiconductor layer in order to provide a potential 
barrier for an electron injected into the semiconductor layer. We will
describe later how one can self-assemble this structure.

\begin{figure}
\epsfxsize=5.3in
\epsfysize=5.8in
\centerline{\epsffile{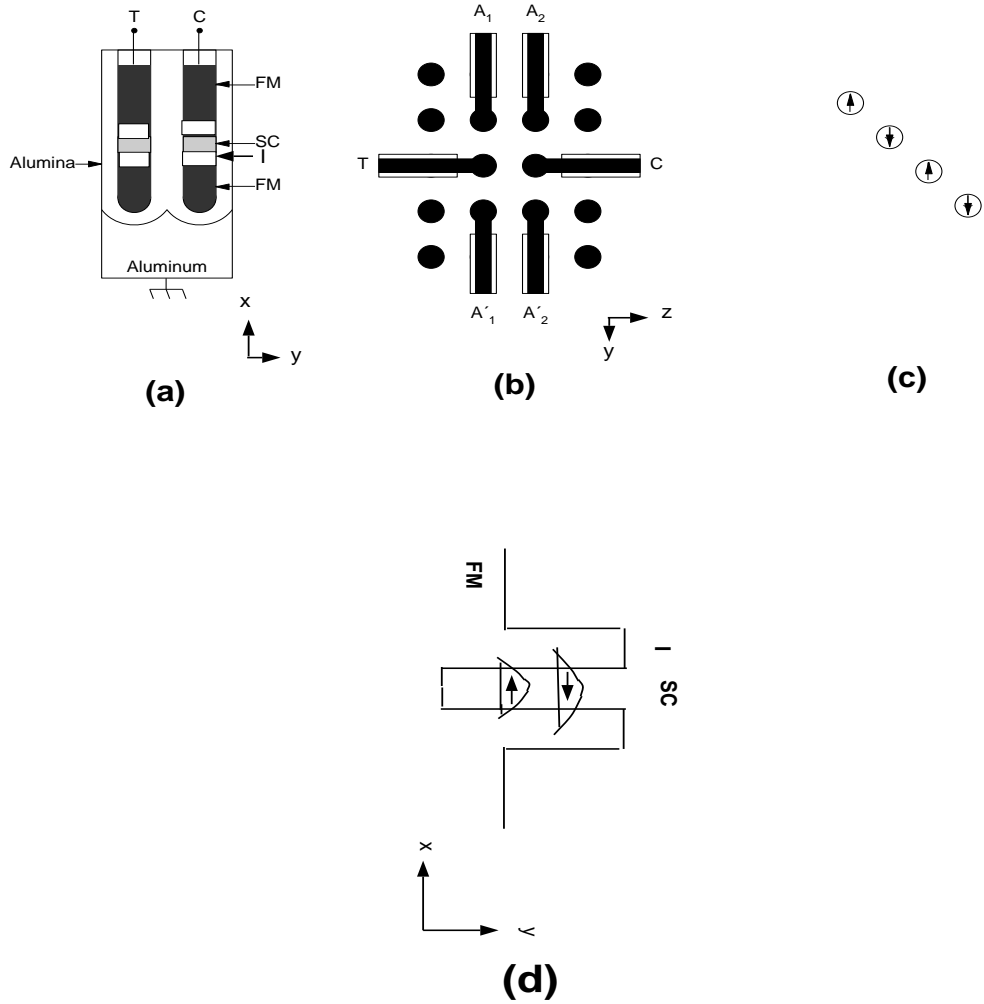}}
\caption[]{\small (a) Cross-section of a penta-layered quantum dot. {\sf FM}
refers to ferromagnet, {\sf SC} refers to semiconductor, and {\sf I} refers
to insulator. (b) The top view showing the ohmic contacts for measuring 
spin polarized current in the target ({\sf T}) and control ({\sf C}) dots,
and the Schottky contacts {\sf $A_1$}, {\sf $A_2$}, {\sf $A'_1$} and {\sf $A'_2$. These contacts are defined by fine-line lithography. (c) a ``spin-wire'' for transmitting quantum information is 
made out of exchange coupled single electron cells.}
which induce the Rashab effect and turn exchange coupling between 
{\sf C} and {\sf T} on and off. (d) The energy band diagram for the 
structure along the x-direction (perpendicular to heterointerfaces) showing the
spin-split states and the corresponding spatial wavefunctions.}
\end{figure}

A dc potential pulse is
applied between the two outer (ferromagnetic) layers to inject a single
spin-polarized electron coherently from one of the outer layers into the middle layer. 
We will discuss the ramifications of such spin injection later. In the
middle layer, the electron's
ground state is spin-split because of Rashba interaction \cite{rashba,rossler,das}. The Rashba
effect arises from spin-orbit coupling in the presence of a transverse
electric field which is always present at the interface of two dissimilar 
materials owing to the 
conduction band discontinuity. It is possible to electrostatically {\it modulate} this 
spin splitting \cite{nitta}
by applying a transverse potential using lithographically delineated 
gate contacts. The applied potential alters the interface field that causes the Rashba effect and hence changes the spin-splitting energy. 

\subsection{Single qubit rotation}

A target qubit is selected for rotation by
bringing its spin-splitting energy in resonance with
the external global ac magnetic field by applying a suitable potential pulse
to the gate contacts.   Arbitrary qubit (spin) rotation is achieved by varying the pulse 
duration, i.e. the duration of resonance with the ac magnetic field. Such a procedure realizes the 
first ingredient of a universal quantum gate, namely
arbitrary single qubit rotations.

\subsection{Two qubit controlled dynamics}

In order to achieve the second and last ingredient of a universal 
quantum gate - namely the conditional dynamics of a universal
2-qubit gate - we need to couple 
the rotation of one qubit (target qubit) with the orientation of another 
qubit (control qubit). This can be done by exploiting the exchange coupling between 
two single electrons in two neighboring dots.
The spin-splitting energy in any dot depends, among other things,
 on the 
spin orientation in the neighboring dot if the two dots are exchange 
coupled.   It is obvious that the total spin splitting $\Delta_{target}$ in
the target dot depends on the exchange interaction
$J$ with the neighboring (control) dot (and hence on the spin orientation of the control qubit) if the two dots are exchange coupled. After all, the exchange term will appear in the 
Hamitonian of the coupled two-dot system. For instance, without 
the exchange interaction, the spin splitting energy in the target dot is
\begin{equation}
E_{\downarrow} - E_{\uparrow} = \Delta
\end{equation}
Now, let us turn the exchange interaction on which lowers the energy of the
singlet state with respect to that of the triplet state. Thus, if the 
control dot's spin is pointing ``up'', the target dot's spin 
will prefer to be ``down''  if the exchange interaction
is operative. This decreases $\Delta$. On the other hand,
if the control dot's spin is pointing ``down'', then $\Delta$ 
is increased. The energy levels are shown in Fig. 2.
Thus, the potential $V_{target}$
that brings the {\it total} spin splitting energy $\Delta_{target}$ in the target dot
in resonance with the ac magnetic field $B_{ac}$ depends on the spin orientation in the control dot. Herein lies the possibility of conditional dynamics.
We can find the $V_{target}$ that will rotate the target qubit through an
arbitrary angle only if the control qubit is in the specified orientation.
Application of this potential $V_{target}$ to the 
target dot realizes the operation of a quantum controlled rotation gate.

To turn the exchange ``on'' and ``off'', one can apply a differential
potential between the target and control dots which will skew the 
wavefunctions and change the overlap between the wavefunctions of
the target and control electrons. Alternately, one can apply positive
potentials to {\it all} Rashba contacts $A_1$, $A_2$, $A'_1$ and $A'_2$.
In this case, there is no differential potential, and the Rashba effect is not
modified, but the positive potential will attract the electrons 
into the insulating layers surrounding the quantum dots, thus increasing 
the overlap between the wavefunctions and the exchange interaction.

\begin{figure}
\epsfxsize=5.8in
\epsfysize=3.4in
\centerline{\epsffile{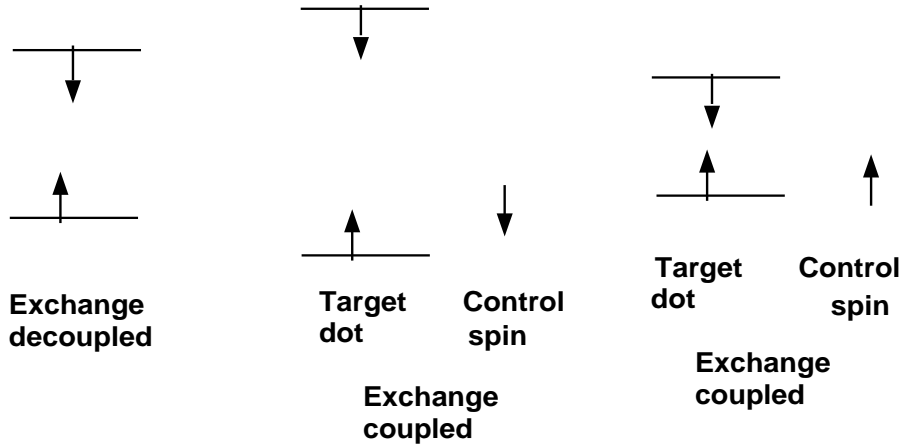}}
\caption[]{\small Energy levels in the target quantum dot depending
on the spin orientation in the control dot.}
\end{figure}

\subsection{Reading the qubit}

Finally, we have to ``read'' a qubit for data output. The qubit (spin orientation in a dot) is read directly via the current
induced between the dot's spin polarized outer layers (ferromagnetic contacts) when a sufficiently
strong potential is applied to overcome the Coulomb blockade. The magnitude
of the current tells us the electron's spin orientation because
 it depends on the angle between the electron's spin
orientation and the direction of magnetization in the ferromagnetic
contact (which is known {\it apriori}). This  principle  was prescribed for measuring 
spin precession in the so-called spin-transistor proposed more than ten 
years ago \cite{das}. It is the same principle that is 
used in tunneling magnetoresistance devices \cite{meservey}.
The single electron current is small, but 
can be measured  using sensitive electrometers. There are alternative
schemes for reading single electron spins that rely on single
electron transistors and the use of the Pauli exclusion principle
\cite{vrijen}. They are much more complicated and are not discussed here.

\section{Rashba effect in a quantum dot}

We will now derive the total spin-splitting in a quantum dot, including the 
Rashba splitting.
Consider the system shown in Fig. 1. The ferromagnetic contacts give 
rise to an in-built magnetic field in the x-direction which can be quite
strong in realistic structures ($\sim$ 1 Tesla). There may be an 
electric field in the x-direction as well to maintain single electron
occupancy and an electric field in the y-direction to induce the
Rashba effect.

The total Hamiltonian for an electron in the semiconductor layer is
\begin{eqnarray}
H  & = & {{({\vec p} - e{\vec A})^2}\over{2m^*}} -e{\cal E}_x x - e {\cal E}_y y
+ (g/2) \mu_B B_x \sigma_x + H_R \nonumber \\
& = & H_0 + H_R
\end{eqnarray}
where $g$ is the Land\'e g-factor, $e$ is the electronic charge,
$\mu_B$ is the Bohr magneton,
$\sigma_x$ is the x-component of the Pauli spin matrix, and $H_R$ is the  Rashba interaction  
given by
\begin{equation}
H_R =  i {{\hbar^2}\over{2m^{*2} c^2}} \nabla V \cdot \left [ {\vec \sigma} \times {\vec \nabla} \right ] = {{ e \hbar}\over{2m^{*2} c^2}} {\vec {\cal E}} \cdot  {\vec \sigma} \times {\vec p}
\end{equation}
where ${\vec \sigma}$ is the Pauli spin matrix, ${\vec p}$ is the 
momentum operator, and ${\vec {\cal E}}$ is the electric field inducing the 
Rashba effect.

We will neglect the effect of the x-directed electric field  on
the Rashba effect and include only the effect of the y-directed field
which is much larger since the potential applied 
along the x-direction must be smaller than $e/2C$ to maintain Coulomb
blockade. In any case, the x-directed field has no Rashba effect
on x-polarized spins, and we are interested in the energy splitting 
between the +x and -x-polarized spins.
Therefore,
\begin{equation}
H_R = {{ e \hbar}\over{2m^{*2} c^2}} {\cal E}_y \left [\sigma_z p_x - \sigma_x p_z \right ]
\end{equation}

The Zeeman term $(g/2) \mu_B B_x \sigma_x$ introduces an Zeeman 
splitting between the +x-polarized spin ($|\uparrow>$) and the 
-x-polarized spin ($|\downarrow>$). If the potential confining
the electron in the semiconductor quantum dot is finite (the 
conduction band offset between the semiconductor and the surrounding
material is finite), then the spatial parts of the ``upspin'' (+x-polarized)
and ``downspin'' (-x-polarized) states are slightly different. The 
wavefunction of the higher energy state will be spread out a little 
bit more. This is shown in Fig. 1(d). Thus, if the downspin state is at a higher energy, then the spatial 
parts of the 
wavefunctions of the two spin states in the lowest spin-split subband of the 
quantum dot can be written as
\begin{eqnarray}
\phi \uparrow = \left ({{2\sqrt{2}}\over{\sqrt{W_x W_y W_z}}} \right )   sin \left ( {{\pi x}\over{W_x}} \right )
  sin \left ( {{\pi y}\over{W_y}} \right ) sin \left ( {{\pi z}\over{W_z}} \right ) \nonumber \\
\phi \downarrow = \left ({{2\sqrt{2}}\over{\sqrt{W'_x W'_y W'_z}}} \right )   sin \left ( {{\pi x}\over{W'_x}} \right )
  sin \left ( {{\pi y}\over{W'_y}} \right ) sin \left ( {{\pi z}\over{W'_z}} \right )
\end{eqnarray}
where $W'_x$ $>$ $W_x$, $W'_y$ $>$ $W_y$ and $W'_z$ $>$ $W_z$. These 
widths are larger than the physical dimensions of the quantum dot
since the wavefunctions will leak out into the barrier as long as the
barrier is not of infinite height. The point of this exercise is to 
show that the spatial parts of the two spin states are different because 
of the Zeeman splitting. This is a critical requirement for the Rashba 
splitting.

Next, we will evaluate the total spin splitting ($\Delta$) which is a
combination of the Zeeman and Rashba splitting. The latter can be modulated by the transverse gate potential. 

The time-independent Schr\"odinger equation describing the ground state of
the system
is 
\begin{equation}
\left (H_0 + H_R \right ) \psi = E \psi
\label{expansion}
\end{equation}

We will expand $\psi$ in the basis functions of the two lowest spin-resolved
subband states. We can neglect the higher subband states as long as the 
Rashba spin splitting $\Delta_R$ is much smaller than the energy 
separation between the lowest two subbands in the quantum dot.
If the effective mass is equal to the free electron mass and the 
dimensions of the quantum dot in all directions is about 10 nm,
then the energy separation between the two lowest subbands is 33 meV.
This is obviously much larger than any reasonable Rashba splitting which scarcely exceeds 1 meV. Hence, neglecting the higher subbands is
justified.

Hence
\begin{eqnarray}
\psi & = &  a_{\uparrow} \phi \uparrow + a_{\downarrow} \phi \downarrow
\end{eqnarray}

Using the above in Equation \ref{expansion}, we get

\begin{equation}
 \left [ \begin{array}{cc}
             <H_1> + <H_R>_{11} & <H_R>_{12} \\
              <H_R>_{21} & <H_2> + <H_R>_{22} \\
             
\end{array}   \right] 
 \left( \begin{array}{c}
             a_{\uparrow} \\
             a_{\downarrow} \\
             
\end{array}   \right)
=
E_{ground}  \left( \begin{array}{c}
             a_{\uparrow} \\
             a_{\downarrow} \\
             \end{array}   \right) ~,
\end{equation}
where $<H_1>$ = $<\phi \uparrow |H_0|\phi \uparrow>$, $<H_2>$ = $<\phi \downarrow |H_0|\phi \downarrow>$, $<H_R>_{11}$ = $<\phi \uparrow |H_R|\phi \uparrow>$, $<H_R>_{22}$ = $<\phi \downarrow |H_R|\phi \downarrow>$,
$<H_R>_{12}$ = $<\phi \uparrow |H_R|\phi \downarrow>$, and $<H_R>_{21}$ = $<\phi \downarrow |H_R|\phi \uparrow>$. 

Diagonalizing the above Hamiltonian, we get
that the total splitting between the upspin and downspin states is
\begin{eqnarray}
E_{\downarrow} - E_{\uparrow} & = & 2 \sqrt{\left ( {{<H_1> - <H_2> + <H_R>_{11} 
- <H_R>_{22}}\over{2}} \right )^2 + <H_R>_{12}<H_R>_{21}  } \nonumber \\
& = & 2 \sqrt{ \left ( {{g \mu_B B}\over{2}} + {{ e \hbar}\over{2m^{*2} c^2}} {\cal E}_y <p_x> \right )^2 + \left |{{e \hbar}\over{2m^{*2} c^2}} {\cal E}_y  <p_z> \right |^2 }
\end{eqnarray}
where $<p_x>$ = $<\phi \uparrow | -i \hbar (\partial/\partial x | \phi \uparrow >$ = $<\phi \downarrow | -i \hbar (\partial/\partial x | \phi \downarrow >$ = 0
and $<p_z>$ = $<\phi \uparrow | -i \hbar (\partial/\partial z | \phi \downarrow >$ = ${{8 i \hbar}\over{\sqrt{W_z W'_z}}}cos^2 \left (\pi  {{ W_x}\over{W'_x}} \right)$.

Therefore the total splitting is
\begin{equation}
\Delta = E_{\downarrow} - E_{\uparrow} = 2 \sqrt{ \left ( {{g \mu_B B}\over{2}}   \right )^2 +  {\cal E}_y^2  {{16 e^2\hbar^4}\over{ m^{*4}c^4 W_x W'_x}} cos^4 \left ({{\pi W_x}\over{W'_x}} \right ) }
\end{equation}

The last term under the radical is the Rashba effect which can be varied
by the electric field ${\cal E}_y$. Note that this term would have 
vanished if $W_x$ = $W'_x$, that is, if the spatial parts of the upspin
and downspin wavefunctions were identical. Here, we have made the 
spatial parts different by using a finite potential barrier and a magnetic 
field to raise the energy of the downspin state above that of the upspin
state.

\subsection{Coherent spin injection from spin polarized contacts}

We mentioned earlier that we wish to inject an electron into the 
quantum dot with a definite spin orientation so that we know
the initial state of the qubit at time $t$ =0.  However,
coherent spin injection from a spin-polarized 
(ferromagnetic) contact into the semiconductor is {\it not} critical for this
purpose.
If the spin-injection does not work, one could still select a definite initial spin using the principle of the spin-RTD.
Since the subbands in the quantum dot are spin split, one can align the 
Fermi level in the contact with one of the spin-split levels and therefore
preferentially inject into that level. This will assign a definite 
spin orientation to the qubit. If even that does not work,
one can inject  an electron with arbitrary spin and wait till the electron decays 
to the lowest state (by a spin flip transition if necessary). Since 
the spin degeneracy is lifted by a combination of the Zeeman and 
Rashba effect (see last equation), the ground state always 
has a definite spin polarization. In the latter two scenarios, ferromagnetic 
contacts are not necessary.

Even though coherent spin ``injection'' is not critical, coherent spin ``detection'' is
absolutely critical since it provides the mechanism for reading a qubit. The injector
is a spin-polarizer and the detector is a spin-analyzer. We can do away
with the polarizer, but not the analyzer. Thus, coherent spin injection across
one of the ferromagnetic interfaces is necessary. If the analyzer does 
not work either, we have to rely on the mechanisms proposed in ref. \cite{vrijen} for reading spin which are much more complicated.

Coherent spin injection from a metal into a semiconductor is a difficult problem
\cite{tang} and has been recently addressed theoretically \cite{schmidt}.
There have been some scattered reports of spin injection from a ferromagnet
into a semiconductor \cite{bennett} and between two semiconductors 
of widely different bandgaps \cite{awschalom}.
Recently, spin polarized hole injection was demonstrated from
GaMnAs into GaAs \cite{ohno} at around a temperature of 120 K. Prior to that, 
spin polarized injection from CdMnTe into CdTe was demonstrated
\cite{oestreich}, but the disadvantage in that case is that CdMnTe is not
a permanent ferromagnet; the
spin polarization needs to be maintained by  a globally applied 
dc magnetic field which introduces a Zeeman splitting in CdMnTe. However,
only a very small field is required since the effective Land\'e g-factor
for electrons in dilute magnetic semiconductors is huge ($\sim$ 100).
On the other hand, the advantage of CdMnTe is that it
is lattice matched to CdTe and hence interface scattering is 
less of a problem.  Most recently,  90\% spin polarized electron injection 
was demonstrated from the dilute magnetic semiconductor Be$_x$Mn$_y$Zn$_{1-x-y}$Se into GaAs 
at a temperature below 5 K \cite{fiederling} and at a relatively
large magnetic field which induces a Zeeman splitting in the magnetic
semiconductor. While the temperature is high enough for quantum
computing applications, the applied magnetic field is too large
and may flip the spin in the semiconductor quantum dot, thus 
corrupting the qubit. The problem of coherent spin injection 
from a ferromagnetic material into a semiconductor is a topic
of much current research. It has a long history and rapid strides are
being made in this field.

Another important question is how easy will it be to maintain 
single electron occupancy in each dot. As long as the energy cost to add an 
additional electron (= $e^2/2C$; $C$ is the capacitance of the dot)
significantly exceeds the thermal energy kT, only a single electron
will occupy each dot. Uniform 
electron occupancy in arrays of $>$ 10$^{8}$ dots has been 
shown experimentally \cite{muerer}.

\subsection{Spin measurement}

After quantum computation is over, we need to read the result by measuring
the qubits. During this process, the qubits will collapse to classical 
bits. These classical bits are the measured spin orientations in relevant
dots. They are measured    
by measuring the current that results when the potential over the 
dot is raised over the Coulomb blockade threshold. If we assume 
that the differential phase-shift suffered by the spin in traversing the
dot is negligible; in other words, transport through the dot does not rotate 
the spin, then the magnitude of the measured current can tell us the
spin orientation \cite{das}. It was shown in ref. \cite{das} that 
the spin-polarized contacts act as electronic analogs
of optical polarizers and analyzers, so the current will depend on the 
projection of
the spin of the quantum dot's resident electron on the spin orientation in the  contacts. Thus, by 
measuring the current, we can tell the spin orientation in any quantum 
dot. 

\subsection{Calibration}

For each dot, the 
potential $V$ that needs to be applied to flip the spin by bringing the 
dot in resonance with $B_{ac}$ can be calibrated following the 
procedure outlined by Kane \cite{kane}. With $B_{ac}$ = 0, we measure 
the spin in a quantum  
dot. Then we switch on $B_{ac}$ and sweep $V$ over a range. Next $B_{ac}$
is switched off and the spin is measured. The range of $V$ is progressively increased
till we find that the spin has flipped. We then proceed to narrow the 
range with successive iteration while making sure that the spin
does flip in each iteration. Finally this allows us to ascertain $V$
with an arbitrary degree of accuracy. As pointed out by Kane \cite{kane}, 
the calibration procedure can, in principle, be carried out in parallel
over several dots simultaneously and the voltages stored in adjacent
capacitors. External circuitry will thus be needed only to control 
the timing of the biases (application of $V_{target}$) and not their
magnitudes. While this is definitely an advantage, fabricating nanoscale capacitors adjacent to each individual 
dot is outside the scope of self-assembly. Moreover, capacitors 
discharge over time, requiring frequent
recalibration through refresh cycles,
so that this may not be a significant advantage.

\subsection{Input and Output Operations}

Any computer is of course useless unless we are able to input and output
data successfully. Since we are using spin-polarized contacts to 
inject an electron in each dot, we know the initial orientation. Those
dots where the initial orientation is the one we want are left unperturbed
while the spins in the remaining dots are flipped by resonating with $B_{ac}$. This process prepares
the quantum computer in the initial state for a computation and can be 
viewed as the act of ``writing'' the input data.
Computation
then proceeds on this initial state by carrying out a desired sequence
of controlled rotation operations. Reading the data is achieved as described
in subsection 2.3.

\subsection{Comparisons with similar proposals}

Proposals similar to that presented in this paper, which envision nanoelectronic spin-based implementations of universal
quantum gates, have been forwarded in the past by Privman \cite{privman}, Kane \cite{kane}
and more recently by  Vrijen, at. al. \cite{vrijen}. Our proposal
is distinct from those previous versions in many ways. The first two of the
previous proposals envision qubits as being encoded by nuclear spins
and a delicate transduction between electron- and nuclear-spins
is required for data communication. Both Vrijen and we have eliminated the
role of nuclear spins (and the need for coupling between electron and 
nuclear spins), but perhaps at the cost of a somewhat smaller T$_2$ time
(spin coherence time).
The major difference between our proposal and all others is that we do not need any dc magnetic field
at all. All previous versions split the spins using
 the Zeeman effect induced by a strong dc magnetic field.
We use the Rashba effect instead (which is purely electrostatic).
Since we only need a small ac magnetic field (supplied by a
microwave source), there is some hope of a ``lightweight'' implementation
where heavy electro-magnets for generating strong dc magnetic fields are not
required. There is nonetheless a 
cryogenic requirement which is the main obstacle to realizing a truly
``portable'' quantum computer. This obstacle is not easy to overcome.

Another major difference with previous proposals is that our structure
can be mostly self-assembled thus eliminating the requirement of 
performing Herculean feats in lithography. In the next section, we briefly
describe how it may be possible to self-assemble a quantum computer.

\section{Self Assembly}

The self-assembly
process that we propose is relatively standard and has been successfully applied by a 
number of groups, including us, for fabricating ordered two-dimensional 
arrays of quantum dots or nanowires \cite{bandy2}. The synthesis proceeds as follows.

First an Al foil is  dc anodized in 15\% sulfuric acid for several hours with a current
density of 40 mA/cm$^{2}$. This produces a nanoporous alumina film
on the surface of the foil with a quasi-ordered arrangement of pores.
The film is stripped off and the foil is re-anodized for a few minutes.
The alumina film that forms on the surface after the second anodization step has a
very well ordered arrangement of nanopores \cite{masuda1}. Fig. 3 shows a raw atomic
force micrograph of pores formed by anodizing in oxalic acid. The pore
diameter is 52 nm and the thickness of the wall separating
two adjacent pores is of the same order. If the anodization is carried
out in sulfuric acid, the pores that self-assemble have a much 
smaller diameter of 10$\pm$1 nm with a wall thickness of the same order
\cite{bandy2}.
Cross-section TEM of the pores have revealed that they are cylindrical
with very uniform diameter along the length. The length of the 
pores is of course the thickness of the alumina film and depends on
the duration of anodization. Typically, the length is a few thousands
of angstroms.

\begin{figure}
\epsfxsize=3.4in
\epsfysize=3.4in
\centerline{\epsffile{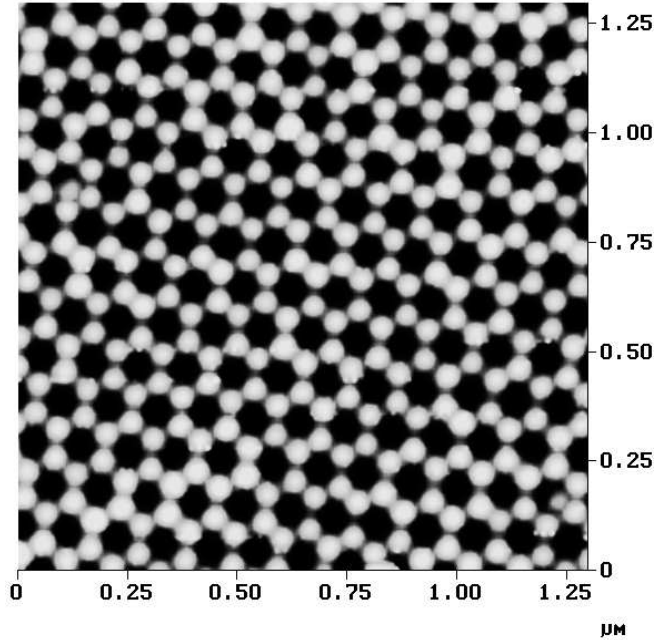}}
\caption[]{\small Raw atomic force micrograph of pore morphologies produced
by anodization of an aluminum foil  in oxalic acid. The average pore 
diameter is 52 nm with a 5\% standard deviation. This structure 
acts as a self-assembled template for self-assembling a quantum 
computer.}
\end{figure}

Multilayered quantum dots as shown in Fig. 1 are formed by sequentially
electrodepositing the constituent layers selectively within the 
pores. However, in order to have appreciable overlap of the 
wavefunctions in neighboring dots for exchange coupling, we must first decrease the thickness
of the alumina walls separating two adjacent dots.  The separation  
can be decreased to as small as $\sim$ 1 nm by widening the pores. This is 
accomplished by soaking the porous alumina film in phosphoric acid
which dissolves the alumina from the walls of the pores.

Electrodeposition of the constituent layers of a multi-layered quantum 
dot (or quantum dash) is carried out in steps. In Fig. 1(a), we show
a penta-layered dot of Co-ZnSe-InAs-ZnSe-Co where Co acts as the spin polarized 
material (it has a 34\% spin polarization). 
For depositing the first Co layer, the alumina film in immersed in a solution of CoSO$_4$
and an ac signal of 20 V rms amplitude and 250 Hz frequency is imposed 
between the aluminum substrate and a graphite counter electrode. During the
cathodic half cycle of the ac signal, the Co$^{++}$ ion is reduced to zero-valent Co metal which goes into pores
selectively since they offer the least impedance path for the ac
current to flow. Since alumina is a valve metal oxide, the zero-valent
Co is not re-oxidized to Co$^{++}$ during the anodic half-cycle. After a few seconds of electrodeposition, we are 
left 
with a $\sim$ 10-nm layer of Co at the bottom of the pore \cite{menon}.

The partially filled alumina film is then ac electrolysed in selenic 
acid for a few seconds which leaves behind the Se$^{--}$ ions adsorbed on 
the walls of the pores. Next, the sample is immersed in a boiling aqueous solution containing the Zn$^{++}$ ion. The Zn$^{++}$ ion reacts with the Se$^{--}$ in the walls 
of the pore to produce a $\sim$ 10-nm thick layer of ZnSe on top of the Fe layer. This material is a wide gap semiconductor and acts as the insulator layer. The pores are then washed in de-ionized water to remove excess 
Se$^{--}$. Next, the narrow gap semiconductor InAs (which has a strong Rashba coupling because of the 
low effective mass), is deposited over the ZnSe layer by carrying out 
ac electrolysis in arsenic acid followed by immersion in a boiling 
solution containing In$^{+++}$ ion. This deposits the InAs layer. 
Finally, ZnSe and Co depositions are repeated.
This results in the structure of Fig. 1(a). Note that the spin-polarized
contacts (Co) are automatically {\it self-aligned} to the semiconductor dot 
(InAs) in this approach.

{\it It should be pointed out that one is not limited by material}.
Almost anything can be deposited selectively within the pores,
one way or another. Even silicon can be deposited by slow deposition
using chemical vapor deposition and Group V elements like carbon have
been deposited within the pores employing essentially gas-phase epitaxy
\cite{moskovitz1}. Plasma-enchanced chemical vapor deposition
is another promising approach.

Material purity is of extreme concern in any electrochemical synthesis.
Chemical reagents are never very pure and we certainly do not want 
a magnetic impurity in the semiconductor dot that will tend to cause
unwanted spin flips. Since it is possible to fill the pores using very 
slow deposition in a CVD set-up, one could use this approach to guarantee vastly improved
material purity with a commensurate increase in fabrication cost. 

\subsection{Wiring the gates to make a computer}

Arbitrary electrical (classical) connections will have to be made between different gates in
order to make a computer. The lithographic challenge associated with 
this task is daunting; however, there is an alternate. We can deposit
Au over the top Co contact in the same way as we deposit Co itself. Gold sulfide
is an appropriate electrochemical source for gold. Conjugated organic 
molecules such as biphenyl dithiol and gold clusters can be co-evaporated
on the surface after each pore is sealed with a top
Au layer. The end-group in the organic molecule self-attaches to 
Au acting as ``alligator-clips'' \cite{tour,henderson}. The molecules 
bridged by Au clusters (Fig. 4a) are electrically conducting with a resistance of 10-40 M$\Omega$ per 
molecule \cite{andres1,andres2}. They are called ``molecular ribbons'' and provide {\it self-assembled}
electrical connection between the quantum dots (Fig. 4a) \cite{datta2}. However, the connection
exists between every dot and hence must be surgically modified to 
realize a specific interconnection pattern. For this purpose, one will need to remove the unwanted
connections  with an STM tip. This is a slow
and laborious process but still beats lithography.

\begin{figure}
\epsfxsize=5.8in
\epsfysize=5.8in
\centerline{\epsffile{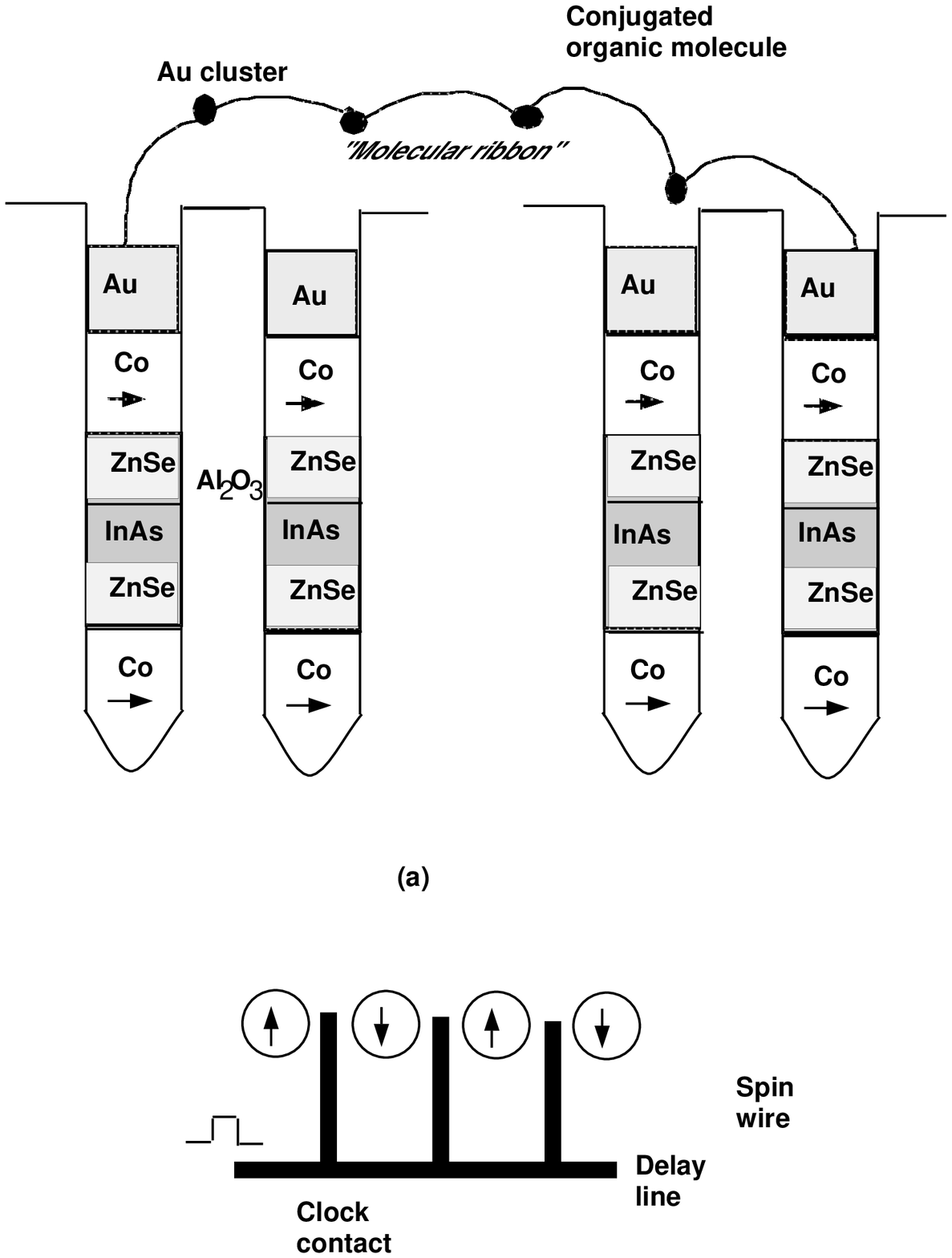}}
\caption[]{\small Wiring the quantum computer. Dot-to-dot connections 
are self-assembled using conjugated organic molecules with appropriate 
end-groups that self-adhere
to gold. Gold clusters act as links in the bridge. Every Au contact 
is connected to others via the linked molecules and the unwanted 
connections are subsequently removed with an STM tip.}
\end{figure}

The electrical interconnects transmit electrical signals such as 
Rashba voltages and spin currents. However, they do not transport
quantum signal such as the state of spin polarization. For this
purpose, we need ``spin wires'' which transmit quantum information about
spin polarization.
Fortunately, making ``spin wires''  is somewhat easier. A line 
of exchange coupled quantum dots is a ``spin wire'' 
since its ground state is anti-ferromagnetic 
\cite{nanotech, molotkov} so that the
spin repeats itself every other dot. This allows spin state to be transmitted
down a line consisting of an odd number of dots. Such a spin wire is
shown in Fig. 4b. Unfortunately such wires are not ``unidirectional''
and do not transmit signal (spin orientation) from the input end
to the output end unidirectionally \cite{nanotech, jjap}. We had 
shown in ref. \cite{jjap} that unidirectionality in time effectively
also imposes unidirectionality in space. Temporal 
(and consequently spatial) unidirectionality can be
 achieved by 
sequential clocking (this is the technique adopted to transmit 
charge packets in conventional charge coupled devices \cite{schroeder}).
We turn the exchange interaction between neighboring dots ``on'' sequentially
by propagating a positive pulse down a line connected to contacts interposed between adjacent
dots as shown in Fig. 4(b).

The above lines will have to be delineated by lithography.
However, since much of the pattern is periodic, it may be possible to 
use techniques such as achromatic interferometric 
x-ray lithography which have a much faster throughput than direct write
lithography to delineate the 
clocking connections. Direct writing is not completely unavoidable,
however. Connections to the 
external world for data input/output  to the entire chip must be delineated
with direct-write lithography. This is however not as demanding as making all the 
internal connections (dot-to-dot connections) with lithography.

\section{Conclusions}

In this paper, we foresee the application of a great advance in materials 
technology, nanoscale self-assembly, to realize a great advance in 
information technology - the quantum computer. Past proposals
of semiconductor
implementations of quantum computers \cite{kane,vrijen} required extremely challenging fabrication methodologies and at least some of them relied on delicate interaction between nuclear and 
electron spins to transduce the qubit into a measurable signal. The present paradigm is much simpler, probably more robust, and the possibility
of self-assembly makes it very attractive.

\section{Acknowledgement}

I am indebted to many individuals for insightful discussion, but especially
to Profs. P. F. Williams and D. J. Sellmyer.

\pagebreak

\end{document}